\documentclass{PoS}

\newcommand{\be}{\begin{equation}}
\newcommand{\ee}{\end{equation}}
\newcommand{\bary}{\begin{eqnarray}}
\newcommand{\eary}{\end{eqnarray}}

\title{Hadronic processes as origin of  TeV emission in Fanaroff-Riley Class I: Cen A, M87 and NGC1275}

\ShortTitle{Fanaroff-Riley Class I}

\author{\speaker{N. Fraija}, M. M. Gonz\'alez and M. P\'erez \\
Instituto de Astronom\'{\i}a, UNAM, M\'exico, 04510,  Universidad Nacional Aut\'onoma de M\'exico, 
Circuito Exterior, C.U., A. Postal 70-264, 04510 M\'exico D.F., M\'exico\\
 \email{nifraija@astro.unam.mx, magda@astro.unam.mx, jguillen@astro.unam.mx}}

\abstract{Recent detections of Fanaroff-Riley Class I  AGNs by HESS, MAGIC, and VERITAS suggest that very-high-energy $\gamma$-rays (VHE, E$\geq$100 GeV) may not have a leptonic origin.  We present a hadronic model to describe the TeV photons as the neutral pion decay resulting from p$\gamma$ and $pp$ interactions. For the p$\gamma$ interaction, we  assume that the target photons  are produced by leptonic processes and apparent at  the second spectral peak. For the pp interaction we consider as targets the thermal particle densities in the  lobes.  We show that this model can describe the TeV spectra of the radio galaxies NCG 1275, M87 and Cen A.}

\FullConference{Gamma-Ray Bursts 2012 Conference -GRB2012,\\
		May 07-11, 2012\\
		Munich, Germany}

\begin{document}


Fanaroff $\&$ Riley Class I (FRI) sources are radio loud active galactic nuclei (AGNs) exhibiting clear structure of a compact central source,  twin  large-scale jets and lobes.  
The spectra of FRI radio galaxies are generally well described by standard non-thermal  Synchrotron and Synchrotron-Self Compton (SSC) models  \cite{tav98,abd10,fra12a}.  The radio through optical emission originates from synchrotron radiation while the  X-ray through $\gamma$-ray emission originates from SSC.  A leptonic model with only one population of electrons  predicts a spectral energy distribution (SED) that can not extend to $\geq 10^{23}$ Hz  \cite{geo05,len08}. Also, some authors have suggested that the GeV to TeV emission may have origins in different physical processes \cite{bro11}. Therefore, we use these two facts and consider hadronic  processes (pp and p$\gamma$) to describe the VHE spectra of  Cen A \cite{abd10,fra12a, aha09},  M87 \cite{abd09b,acc10} and NGC1275\cite{abd09a, ale10}.  We first require a description of the SED up to MeV/GeV   \cite{abd10,fra12a, abd09b, abd09a}. 
Then, we assume the thermal particle densities in the  lobes  and the lobe's distance to the core \cite{har09, fra12b} to describe the TeV energy contribution to the spectrum either by p$\gamma$ or pp interaction.

NGC 1275 is located at the centre of the Perseus cluster at a redshift of $z=0.0179$.  It has a strong, compact nucleus and jet.  Its jet has been  detected with an intrinsic velocity of $0.3c - 0.5c$ and oriented at an angle of $\approx$ 30$^\circ$ - 55$^\circ$ to the line of sight \cite{abd09a}.  M87 is located in the Virgo cluster of galaxies at a distance of $\sim$ 16 Mpc (z=0.0043).  It has been detected from radio to VHE gamma rays. In particular,  it was detected  by HESS, MAGIC and VERITAS showing a variable TeV emission on timescales of years  \cite{abd09b}, although much faster variations, down to timescales of a day and less, have been observed \cite{abr12, acc09}.    Cen A is at a distance of 3.8 Mpc.  Its giant radio lobes, which subtend $\sim\,10^\circ$ on the sky, are oriented primarily in the north-south direction. Also,  Cen A has a  jet with an axis subtending an angle to the line of sight estimated as $15^\circ\,-\,80^\circ$ \cite{abd10, fra12a}.
 
 \section{Hadronic interactions }

We assume for protons a power law injection spectrum given by  $dN_p/dE_p=A_p\,E_p^{-\alpha_p}$, where $\alpha_p$ is the spectral index and $A_p$ is the proportionality constant.  

We assume that the p$\gamma$ interaction takes place when accelerated protons collide with target photons \cite{ato03} originated in the SSC process. Then, the energy loss rate due to pion production is given by $t'_{p,\gamma}=1/(2\,\gamma_p)\int^\infty_{\epsilon_0}\,d\epsilon\,\delta_\pi(\epsilon)\xi(\epsilon)\,\epsilon\int^\infty_{\epsilon/2\gamma_p}dx\, x^{-2}\,n(x)$ \cite{ste68}, where $n(x)=dn_\gamma/d\epsilon_\gamma (\epsilon_\gamma=x)$, $\sigma_\pi(\epsilon)$ is the cross section for pion production for a photon with energy $\epsilon$ in the proton rest frame, $\xi(\epsilon)$ is the average fraction of energy transfered to the pion,  $\epsilon_0=0.15$ is the threshold energy and, $\gamma_p=\epsilon_p/m^2_p$. The fraction of energy lost is $f_{\pi^0,p \gamma}\approx t'_d/t'_{p,\gamma}$,  (where $t'_d\sim r_d/\Gamma$ is the expansion time scale)  and the differential spectrum, $dN_\gamma/dE_\gamma$, of the photon-pions produced by  p$\gamma$ interaction  is given by, 

\begin{small}
\begin{equation}
\label{pgamma}
\left(E^2\,\frac{dN}{dE}\right)^{obs}_{\pi^0-\gamma} = A_{p,\gamma}
\cases{
\left(\frac{E^{obs}_{\pi^0-\gamma,c}}{E_{0}}\right)^{-1} \left(\frac{E^{obs}_{\gamma}}{E_{0}}\right)^{-\alpha_p+3}          &   $E^{obs}_{\gamma} < E^{obs}_{\pi^0-\gamma,c}$\cr
\left(\frac{E^{obs}_{\gamma}}{E_{0}}\right)^{-\alpha_p+2}                                                                                        &   $E^{obs}_{\pi^0-\gamma,c} < E^{obs}_{\gamma}$\cr
}
\end{equation}
\end{small}

\noindent where
\begin{small}
\be
A_{p,\gamma}\propto \delta_D^{\alpha_p} \,E_0^2\,A_p\,  e^{-\tau_{\gamma\gamma}}   (1+z)^{-\alpha_p} \,{E^{obs}_{\gamma,c}}^{-1}\, L^{obs}\,dt^{obs}\,{d_z}^{-2}
\ee
\end{small}

\noindent and
\begin{small}
\be
E^{obs}_{\pi^0-\gamma,c}= 5\times 10^{-2}\,\frac{ \delta_D^2\,(m_\Delta^2-m_p^2)}{(1+z)^2(1-cos\theta)} {E^{obs}_{\gamma,c}}^{-1}
\ee
\end{small}

\noindent where $L^{obs}$ is the observed luminosity,  $dt^{obs}$  is  the observed variability,  $\tau_{\gamma\gamma}$ is the optical depth \cite{fra12a},   $\delta_D$ is the  Doppler factor and $ d_z$ is the distance to the source.

On the other hand, for pp interactions we assume that  the accelerated protons collide with thermal particles in the giants lobes. Then, the energy loss rate due to pion production is given by  $ t'_{pp}=(n'_p\,k_{pp}\,\sigma_{pp})^{-1} $\cite{ato03}, where $\sigma_{pp}=30$ mbarn is the nuclear interaction cross section, $k_{pp}=0.5$ is the inelasticity coefficient and $n'_p$ is the comoving thermal particle density.  The fraction of energy lost by pp is  $f_{\pi^0,pp}\approx t'_d/t'_{pp}$ and the differential spectrum of the photon-pions produced by  pp interaction, $dN_\gamma/dE_\gamma$,   is given by, 
\begin{small}
\begin{equation}
\label{pp}
\left(E^{2}\, \frac{dN}{dE}\right)^{obs}_{pp,\gamma}= A_{pp}\, \left(\frac{E^{obs}_{\gamma}}{E_{0}}\right)^{2-\alpha_p}
\end{equation}
\end{small}
with,
\begin{small}
\be
A_{pp}\propto \frac{\Gamma^2\,\delta_D^{2+\alpha_p}\,E_0^2\,A_p\,e^{-\tau_{\gamma\gamma}}}  {(1+z)^{2+\alpha_p}}\,R\,n_p\,{dt^{obs}}^2\, d_z^{-2}
\ee
\end{small}

\noindent where $n_p$ is the thermal particle density,  $R$ is the distance to the lobes from the AGN core and $\Gamma$ is the bulk Lorentz factor. The eqs. \ref{pgamma} and \ref{pp}  represent the TeV energy range contribution  to the spectrum. We have considered only one hadronic interaction at once.

The input parameters used to fit the leptonic contribution to the SED of  Cen A \cite{abd10,fra12a},  M87 \cite{abd09b} and NGC1275\cite{abd09b} are shown in Table 1. In Table 2 the parameters used and their values to describe  the TeV  contribution  to the spectrum  considering p$\gamma$ or pp interactions are shown.

\begin{center}\renewcommand{\arraystretch}{0.75}\addtolength{\tabcolsep}{-1pt}
\begin{tabular}{ l c c c }
 \hline

\hline
\hline
\normalsize{Parameters} & \normalsize{Cen A} & \normalsize{M87} &  \normalsize{NGC1275} \\
\hline
\hline

\scriptsize{$dt^{obs}$ (s)} & \scriptsize{$2.5 \times 10^{6}\,$}  & \scriptsize{$1.0 \times 10^{5}\,$}  & \scriptsize{$3.17 \times 10^{7}\,$}   \\
 \scriptsize{$L^{obs}$ (erg s$^{-1}$) } & \scriptsize{$5 \times 10^{43}\,$} &\scriptsize{$1.0 \times 10^{44}\,$} & \scriptsize{$6 \times 10^{43}\,$}  \\ 
\scriptsize{ $\theta$ (degrees)} & \scriptsize{$40\,$}  &  \scriptsize{$10\,$}  & \scriptsize{$20\,$}   \\
\scriptsize{$\delta_{d}$} & \scriptsize{1.47} &  \scriptsize{3.9} &  \scriptsize{2.9} \\
\scriptsize{$d_z$ (pc)} & \scriptsize{$3.8$}&\scriptsize{$16$}&\scriptsize{$76$} \\

\hline
\hline

 \end{tabular}
\end{center}

\begin{center}
\scriptsize{\textbf{Table 1. Input parameters used to fit the leptonic contribution of the SED}}\\
\scriptsize{}
\end{center}


\begin{center}\renewcommand{\arraystretch}{0.75}\addtolength{\tabcolsep}{-1pt}
\begin{tabular}{ |l c c c |l c c c |}

\hline
\hline
\normalsize{Parameters} & \normalsize{Cen A} & \normalsize{M87} &  \normalsize{NGC1275} & \normalsize{Parameters} & \normalsize{Cen A} & \normalsize{M87} &  \normalsize{NGC1275}  \\
\hline
\hline

\scriptsize{$\alpha_p$} & \scriptsize{2.82}  & \scriptsize{2.49}  & \scriptsize{2.7}    &  \scriptsize{$\alpha_p$} & \scriptsize{2.82}  & \scriptsize{2.49}  & \scriptsize{2.7}  \\
\scriptsize{E$^{obs}_{\pi^0-\gamma,c}$ (GeV)} & \scriptsize{317.1} &\scriptsize{674.9} &\scriptsize{80,6} &  \scriptsize{$n_p (cm^{-3})$} & \scriptsize{$10^{-4}$}  & \scriptsize{$10^{-1}$}  & \scriptsize{$10^{-1}$}  \\
\scriptsize{A$_{p\gamma}$  (MeV cm$^2$ s)$^{-1}$ }   & \scriptsize{1.37$\times10^{-2}$}  & \scriptsize{5.37}  &\scriptsize{4.37$\times10^{2}$} & \scriptsize{A$_{pp}$  (MeV cm$^2$ s)$^{-1}$ }   & \scriptsize{ 5.9$\times10^{-13}$}   & \scriptsize{ 2.0 $\times10^{-14}$ } &  \scriptsize{7.2 $\times10^{-14}$ }    \\
\scriptsize{E$_0$ (TeV)}   & \scriptsize{1 } &\scriptsize{1}  & \scriptsize{1 } & \scriptsize{E$_0$ (TeV)}   & \scriptsize{1 } &\scriptsize{1}  & \scriptsize{1 }    \\
                                              &			 &			&			 & \scriptsize{R (kpc)}	     & \scriptsize{100 } &  \scriptsize{150}    &  \scriptsize{200 }\\

\hline
\hline

 \end{tabular}
\end{center}

\begin{center}
\scriptsize{\textbf{Table 2. Parameters used to describe  the TeV  spectrum  considering p$\gamma$ (left) or pp (right) interactions are shown. \cite{fra12b}}}\\
\scriptsize{ }
\end{center}

\section{Discussion  and Conclusions}

When p$\gamma$ interaction is considered and assuming that each photon-pion  carries $\sim $ 10$\%$,  we obtain a broken-power law (eq. \ref{pgamma}) with  break energies inside the energy range  and spectral indexes observed by HESS,  VERITAS and MAGIC detections of Cen A,  M87 and NGC1275 respectively. When pp interaction is considered and assuming that each pion carries $\sim $ 18$\%$ and splits in two $\gamma-$rays, we also obtain energy  ranges observed by the HESS,   VERITAS and MAGIC. For Cen A the  number density of thermal particles in the giants lobes and the distance lobes-core are known, Hardcastle et al. (2009). However  for the M87 and NGC1275,  these values are not well-determined.  Fraija et al. (2012)  shows  the proportionality constant (App) as a function of  distance to the core (R) for typical number density of thermal particles ($n_p$), obtaining the minimum App values for this process. 

We observe that p$\gamma$ interaction is much stronger that pp interaction. For Cen A, the extrapolation to ultra-high energies (UHE) of the proton spectrum could explain the observed UHE cosmic rays (UHECR) observed by the Auger Observatory \cite{fra12a} only for case when pp interactions are the responsibles of the TeV emission . Otherwise the expected number of UHECR obtained is several orders of magnitude above the observations.  We have taken into account the variability for M87, $\sim  10^{5}$ s, that corresponds to low state, for short  timescale variations presented in flaring state  this model  cannot be accommodated.

\end{document}